\documentclass[preprintnumbers,prd,showpacs,floatfix,superscriptaddress,nofootinbib,twocolumn, letterpaper]{revtex4-1}
\pdfoutput=1

\usepackage{longtable}
\usepackage{bm}
\usepackage{relsize}
\usepackage{amsfonts}
\usepackage{amsmath}
\usepackage{amssymb,epsf}
\usepackage{latexsym}
\usepackage{graphicx,epsfig}
\usepackage{amssymb}
\usepackage{float}
\usepackage{subfigure}
\usepackage{epstopdf}
\usepackage[colorlinks=true,citecolor=blue,linkcolor=blue,urlcolor=black]{hyperref}
\usepackage{dcolumn}
\usepackage{psfrag}
\usepackage{wrapfig}
\usepackage{makeidx}
\usepackage{epsf}
\usepackage{color}
\usepackage{multirow}
\usepackage{mathtools}

\begin{document}

\title{Accretion disks and relativistic line broadening in boson star spacetimes}

\author{João Luís Rosa}
\email{joaoluis92@gmail.com}
\affiliation{Institute of Physics, University of Tartu, W. Ostwaldi 1, 50411 Tartu, Estonia}
\affiliation{Institute of Theoretical Physics and Astrophysics, University of Gda\'{n}sk, Jana Ba\.{z}y\'{n}skiego 8, 80-309 Gda\'{n}sk, Poland}

\author{Joaqu\'in Pelle}
\email{pelle.joaquin@gmail.com}
\affiliation{Instituto de F\'isica Enrique Gaviola, CONICET, Ciudad Universitaria, 5000 C\'ordoba, Argentina}
\affiliation{Facultad de Matem\'atica, Astronom\'ia, F\'isica y Computaci\'on, Universidad Nacional de C\'ordoba, Argentina}

\author{Daniela P\'erez}
\email{danielaperez@iar.unlp.edu.ar}
\affiliation{Instituto Argentino de Radioastronom\'ia (IAR, CONICET/CIC/UNLP), C.C.5, (1894) Villa Elisa, Buenos Aires, Argentina} 

\date{\today}

\begin{abstract} 

In this work, we analyze the observational properties of static, spherically symmetric boson stars with fourth and sixth-order self-interactions, using the Julia-based general-relativistic radiative transfer code Skylight. We assume the boson stars are surrounded by an optically thick, geometrically thin accretion disk. We use the Novikov-Thorne model to compute the energy flux, introducing a physically based accretion model around these boson star configurations.  Additionally, we calculate the relativistic broadening of emission lines, incorporating a lamppost corona model with full relativistic effects for the first time around a boson star.
Our results show distinct observational features between quartic-potential boson stars and Schwarzschild black holes, owing to the presence of stable circular orbits at all radii around the former. On the other hand, compact solitonic boson stars, which possess an innermost stable circular orbit, have observational features closely similar to black holes. This similarity emphasizes their potential as black-hole mimickers. However, the compact boson stars, lacking an event horizon, have complex light-ring structures that produce potentially observable differences from black holes with future generations of experiments.

\end{abstract}


\maketitle

\section{Introduction}\label{sec:intro}

In recent years, the rapid development of the experimental component of gravitational physics has successfully provided evidence for the existence of extremely compact objects. In particular, the detection of gravitational waves from binary mergers by the LIGO-Virgo Collaboration (LVC) \cite{LIGOScientific:2016aoc,KAGRA:2021vkt}, the observation of a shadow in the core of the M87 galaxy \cite{EventHorizonTelescope:2019dse,EventHorizonTelescope:2021bee} and near Sgr A* \cite{EventHorizonTelescope:2022wkp} by the EHT collaboration, along with the observation of infrared flares close to the galactic centre by the GRAVITY collaboration \cite{GRAVITY:2020lpa,GRAVITY:2023avo}, compose a set of observations consistent with the theoretical predictions of black hole (BH) spacetimes. Indeed, these observations stand within a close agreement with the Kerr hypothesis \cite{Will:2014kxa,Yagi:2016jml}, describing the final state of a complete gravitational collapse in an astrophysical setting as a rotating electrically-neutral BH \cite{Kerr:1963ud,Penrose:1964wq}.

Despite the apparent success of BH spacetimes in explaining the observations above, it is well known that BH spacetime models are intrinsically problematic from a mathematical and physical point of view. BH spacetimes models in general relativity are singular \cite{Penrose:1964wq,Penrose:1969pc}, i.e., under certain assumptions, the solutions of the field equations are defective beyond repair, thus pointing out the incompleteness of the theory \cite{Romero:2013ag}. On the other hand, the existence of an event horizon seems to violate the unitary evolution required by Quantum Mechanics \cite{Hawking:1976ra}. To overcome these limitations, a wide plethora of alternative models known as exotic compact objects (ECOs) has been developed (see \cite{Cardoso:2019rvt} and references therein), to reproduce the same observational properties of BH spacetimes (in which case they are called BH mimickers) without featuring the same fundamental problems.

Among the models proposed for ECOs, one of the most popular are the self-gravitating condensates of bosonic fields, commonly known as boson stars \cite{Visinelli:2021uve}. The main advantage of such models in comparison to other alternatives stands on the fact that dynamical mechanisms for the formation of such condensates are known \cite{Liebling:2012fv,Brito:2015yga,Brito:2015yfh}. Furthermore, this framework allows for the development of a wide variety of models composed of fundamental fields with different spins and interaction potentials \cite{Brito:2015pxa,Cardoso:2021ehg}, with consequent implications in several phenomenological contexts e.g. X-ray spectroscopy \cite{Cao:2016zbh,Dove:1997ei}, dark matter \cite{Hui:2016ltb}, and gravitational waves \cite{Palenzuela:2017kcg}, the latter due to the rise of additional observational signatures as, e.g., gravitational echoes \cite{Cardoso:2016oxy,Cardoso:2017cqb,Cardoso:2016rao} and tidal effects \cite{Postnikov:2010yn,Cardoso:2017cfl}. Given their physical relevance, an active effort to test the observational properties of these ECOs is currently undergoing \cite{Herdeiro:2021lwl,Rosa:2022tfv,Rosa:2022toh,Rosa:2023qcv,Olivares:2018abq,Gjorgjieski:2023qpv}. 

This work aims to extend the existing literature by introducing physically principled radiative models that have not been considered before. Previously, observational features of thin accretion disks with ad-hoc intensity profiles have been studied around the boson star configurations that we adopt \cite{Rosa:2022tfv,Rosa:2022toh,Rosa:2023qcv}. Here, we apply the Novikov-Thorne model to calculate temperature profiles, thus introducing a physically principled accretion model around these configurations for the first time. On the other hand, we calculate the relativistic broadening of spectral lines. Similar calculations have been done before \cite{lu2003, cao2016}, using ad-hoc emissivity profiles such as broken power laws and lamppost corona prescriptions from flat spacetime. Here, we incorporate a lamppost corona model while taking into account the full relativistic effects on the illumination profile.

In this work, we use \texttt{skylight}\footnote{github.com/joaquinpelle/Skylight.jl} \cite{Pelle:2022sky}, a modern open-source Julia-based \cite{Bezanson:2017Julia} general-relativistic radiative transfer code, to calculate the predicted images and spectra associated with the radiative models we introduce. This code was previously used to produce observational properties of millisecond pulsars \cite{Carrasco:2022kgk} and charged traversable wormholes \cite{Neto:2022pmu}. Given the versatility of \texttt{skylight} to support custom spacetimes, in this work we incorporate the boson star models previously considered in Ref.~\cite{Rosa:2023qcv} to the code.     

This manuscript is organized as follows. In Sec.~\ref{sec:theory} we introduce the Einstein-Klein-Gordon theory and briefly review how to obtain the bosonic star configurations used in the subsequent sections; in Sec.~\ref{sec:disk} we introduce the Novikov-Thorne model for the accretion disk and compute the thermodynamical properties of the disks in the boson star configurations considered; in Sec.~\ref{sec:results} we use \texttt{skylight} to calculate the observed intensities of the accretion disks and the relativistic broadening of spectral lines; finally, in Sec.~\ref{sec:concl} we trace our conclusions and discuss possible pathways for future work.

\section{Theory and framework}\label{sec:theory}

In this work, we are interested in studying self-gravitating scalar field configurations known as bosonic stars. These models are solutions of the Einstein-Klein-Gordon (EKG) system of equations. The EKG theory is described by an action $S$ given by
\begin{equation}\label{eq:action}
    S=\int_\Omega\sqrt{-g}\left[\frac{R}{2\kappa^2}-\frac{1}{2}\partial_\mu\Phi\partial^\mu\Phi^*-\frac{1}{2}V\left(|\Phi|^2\right)\right]d^4x,
\end{equation}
where $\Omega$ is the spacetime manifold on which a set of coordinates $x^\mu$ is defined, $g$ is the determinant of the metric $g_{\mu\nu}$, $R=g^{\mu\nu}R_{\mu\nu}$ is the Ricci scalar, with $R_{\mu\nu}$ the Ricci tensor, $\kappa^2=8\pi G/c^4$ where $G$ is the gravitational constant and $c$ the speed of light, $\partial_\mu$ denote partial derivatives, $\Phi$ is a complex scalar field, $\ ^*$ denotes complex conjugation, and $V\left(|\Phi|^2\right)$ is a potential function depending solely on $|\Phi|^2=\Phi^*\Phi$. In what follows, we adopt a system of geometrized units in such a way that $G=c=1$, and consequently $\kappa^2=8\pi$.

The Einstein-Klein-Gordon system of equations of motion can be obtained via the variational method approach to Eq.~\eqref{eq:action} by taking a variation with respect to the metric $g_{\mu\nu}$ and the scalar field $\Phi$. This system takes the form
\begin{equation}\label{eq:field}
G_{\mu\nu}=8\pi T_{\mu\nu},
\end{equation}
\begin{equation}\label{eq:eom_phi}
\left[\Box-V'\left(|\Phi|^2\right)\right]\Phi=0,
\end{equation}
where $G_{\mu\nu}=R_{\mu\nu}-\frac{1}{2}g_{\mu\nu}R$ is the Einstein's tensor, $\Box=g^{\mu\nu}\nabla_\mu\nabla_\nu$ is the d'Alembert operator with $\nabla_\mu$ denoting the covariant derivatives, a prime denotes a derivative of a function with respect to its sole argument, and $T_{\mu\nu}$ is the stress-energy tensor of the scalar field given explicitly by
\begin{equation}\label{eq:tmunu}
T_{\mu\nu}=\nabla_{(\mu}\Phi^*\nabla_{\nu)}\Phi-\frac{1}{2}g_{\mu\nu}\left(\nabla_\sigma\Phi^*\nabla^\sigma \Phi+V\right),
\end{equation}
where we have introduced index symmetrization $X_{(\mu\nu)}=\frac{1}{2}\left(X_{\mu\nu}+X_{\nu\mu}\right)$.

Static and spherically symmetric bosonic star configurations can be obtained as solutions of the system above by considering the following ansatz for the metric $g_{\mu\nu}$ and for the scalar field $\Phi$ in the usual set of spherical coordinates $x^\mu=\left(t,r,\theta,\varphi\right)$ as
\begin{equation}\label{eq:metric}
    ds^2=-A\left(r\right)dt^2+\frac{1}{B\left(r\right)}dr^2+r^2\left(d\theta^2+\sin^2\theta d\varphi^2\right),
\end{equation}
\begin{equation}\label{eq:phi}
    \Phi=\phi\left(r\right)e^{-i\omega t},
\end{equation}
where $A\left(r\right)$ and $B\left(r\right)$ are metric functions, $\phi\left(r\right)$ is the radial wave function of the scalar field, and $\omega$ is the frequency of the scalar field. Introducing Eqs.~\eqref{eq:metric} and \eqref{eq:phi} into Eqs.~\eqref{eq:field} and \eqref{eq:eom_phi} leads to the following system of equations
\begin{equation}\label{eq:system1}
    \frac{B'}{r}+\frac{B-1}{r^2}=-2\pi\left(\frac{\omega^2\phi^2}{A}+B\phi'^2+V\right),
\end{equation}
\begin{equation}\label{eq:system2}
    \frac{BA'}{rA}+\frac{B-1}{r^2}=2\pi\left(\frac{\omega^2\phi^2}{A}+B\phi'^2\right),
\end{equation}
\begin{equation}\label{eq:system3}
    \frac{1}{2}\phi'\left(\frac{BA'}{A}+\frac{4B}{r}+B'\right)+\phi\left(\frac{\omega^2}{A}-V'\left(|\Phi|^2\right)\right)+B\phi''=0.
\end{equation}
Due to its complexity, the system of Eqs.~\eqref{eq:system1} to \eqref{eq:system3} does not allow for analytical solutions and must instead be solved numerically. A common method used to solve this system is the shooting method for the parameter $\omega$ subjected to suitable boundary conditions at the origin and at infinity that preserve both the regularity (absence of divergences) and asymptotic flatness of the solutions. These boundary conditions are
\begin{eqnarray}
    &&A\left(r\to\infty\right)=1-\frac{2M}{r},\\
    &&B\left(r\to\infty\right)=1-\frac{2M}{r},\\
    &&\phi\left(r\to\infty\right)=1,
\end{eqnarray}
at infinity, where $M$ is a constant denoting the total mass of the boson star, and
\begin{eqnarray}
    &&A\left(r\to 0\right)=A_0,\\
    &&B\left(r\to 0\right)=1,\\
    &&\phi\left(r\to 0\right)=\phi_0,\\
    &&\phi'\left(r\to 0\right)=0,
\end{eqnarray}
where $A_0$ and $\phi_0$ are free parameters. Note that the parameter $A_0$ can always be rescaled to $1$ via a time reparametrization that preserves the asymptotic behavior. Furthermore, in order to solve the system of Eqs.~\eqref{eq:system1} to \eqref{eq:system3}, one needs to specify an explicit form of the potential $V\left(|\Phi|^2\right)$. Depending on the form of this potential, different types of bosonic star configurations arise. Common forms of the potential are the following
\begin{equation}\label{eq:pot1}
V\left(|\Phi|^2\right)=\mu^2|\Phi|^2,
\end{equation}
\begin{equation}\label{eq:pot2}
V\left(|\Phi|^2\right)=\mu^2|\Phi|^2+\Lambda |\Phi|^4,
\end{equation}
\begin{equation}\label{eq:pot3}
V\left(|\Phi|^2\right)=\mu^2|\Phi|^2\left(1+\frac{|\Phi|^2}{\alpha^2}\right)^2,
\end{equation}
where $\mu$ is a constant parameter playing the role of the mass of the scalar field $\Phi$, $\Lambda$ is a coupling constant measuring the intensity of the quartic self-interaction of $\Phi$, and $\alpha$ is an additional coupling constant associated with the sixth-order self-interactions of $\Phi$. Bosonic star configurations under the assumption of Eq.~\eqref{eq:pot1} are commonly denoted as mini-boson stars, configurations obtained under the assumption of Eq.~\eqref{eq:pot2} are known as massive or $\Lambda$-boson stars, and configurations arising from the assumption of Eq.~\eqref{eq:pot3} are known as solitonic boson stars. In previous works \cite{Rosa:2022tfv,Rosa:2022toh,Rosa:2023qcv}, it was shown that the observational properties of mini and $\Lambda$-boson stars are qualitatively similar, while the latter can feature solutions with larger masses. Thus, for this paper, we chose to work solely with $\Lambda$-boson stars and solitonic boson stars, i.e., the potentials take the forms of Eqs.~\eqref{eq:pot2} and \eqref{eq:pot3}, respectively. Furthermore, to provide a comparison ground with previous publications, we choose to work with the same set of configurations as previously analyzed in the context of the orbital motion of hot spots and optically-thin accretion disks. These solutions are summarized in Table \ref{tab:solutions}, which were obtained for $\Lambda=400$ and $\alpha=0.08$, and are linearly stable against radial perturbations. For a boson star of mass $M$, and with a given radius $R$ encompassing $98\%$ of the total mass, the compacticity of the star is defined as
\begin{equation}
\mathcal C = \frac{M}{R}  
\end{equation} 
where $\mathcal C$ depends on the specific boson star model (see Table \ref{tab:solutions}). In Table \ref{table:1}, we specify the value of the radius of the boson stars as defined above and the radius of the ISCO for the solitonic models, both normalized to the gravitational radius $r_g=GM/c^2$. For additional details on the configurations considered, we refer the reader to Ref. \cite{Rosa:2023qcv}. In the following Section, we analyze the accretion process for each of the configurations considered.

\begin{table}
	\caption{Bosonic star configurations considered in this work and the values of their respective parameters $\phi_c$, $\mu M$, $\mu R$, $\mathcal C$, and $\omega/\mu$, where $\mathcal C\equiv M/R$ denotes the compacticity and $R$ denotes the radius encapsulating $98\%$ of the total mass of the star. $\Lambda$BS models correspond to the potential in Eq.~\eqref{eq:pot2}, whereas SBS models correspond to the potential in Eq.~\eqref{eq:pot3}.}
    \label{tab:solutions}
	\begin{tabular}{|c||c c c c c|}
		\hline
		Label & $\phi_c$ & $\mu M$ & $\mu R$  & $\mathcal{C}$ & $\omega/\mu$ \\
		\hline
		 $\Lambda$BS1 & 0.03045  & 1.6321 & 16.1577 & 0.10101 & 0.88124 \\  
		 $\Lambda$BS2 & 0.03457 & 1.7356 & 14.9648 & 0.11597 & 0.86410\\
		 $\Lambda$BS3 & 0.04582 & 1.8368 & 12.4524 & 0.14750 & 0.82786\\
		 SBS1 & 0.0827 &  1.7531 & 11.5430 & 0.1518 & 0.25827 \\
		 SBS2 & 0.0827 & 4.220 &  16.6520 & 0.25342 & 0.17255 \\
		 SBS3 & 0.0850 &  5.655 & 17.6470 & 0.32045 & 0.13967 \\
		\hline
	\end{tabular}
\end{table}

\section{Accretion disks around boson stars}\label{sec:disk}


We model the accretion process onto the boson star by considering a relativistic geometrically thin accretion disk in a quasi-steady state (i.e. such that the mass accretion rate $\dot{M}$ remains constant over time). Novikov and Thorne \citep{Novikov:1973kta} and Page and Thorne \citep{Page:1974he} derived an expression for the heat $Q$ emitted by a relativistic accretion disk in a stationary, axisymmetric spacetime. The expression for the flux $Q$ can be written in terms of the specific energy $\tilde{E}$, the specific angular momentum $\tilde{L}$, and the angular velocity $\Omega$ of the particles that move on equatorial circular geodesic orbits around the central object, which is given by
\begin{equation}\label{flux}
Q(r) = - \frac{\dot{M}}{4 \pi \sqrt{- g}} \frac{\partial_r\Omega}{\left(\tilde{E} - \Omega \tilde{L}\right)^{2}} \int^{r}_{r_{\text{ISCO}}}  \left(\tilde{E} - \Omega \tilde{L}\right) \partial_r\tilde{L} dr,
\end{equation}
where $g$ denotes the metric determinant, $\dot{M}$ is the mass accretion rate, and $r_{\text{ISCO}}$ stands for  the radius of the ISCO. As shown by Harko et al. \citep{Harko:2009xf}, the quantities $\tilde{E}$, $\tilde{L}$ and $\Omega$ depend only on the metric coefficients  $g_{tt}$, $g_{t\phi}$ and $g_{\phi\phi}$, and on the derivatives of these coefficients with respect to the radial coordinate. In a static and spherically symmetric spacetime, $\tilde{E}$, $\tilde{L}$ and $\Omega$ take the forms
\begin{eqnarray}
\tilde{E} & = & - \frac{g_{tt}}{\sqrt{- g_{tt} - g_{\phi\phi} \Omega^2}},\\
\tilde{L} & = & \frac{g_{\phi\phi} \Omega}{\sqrt{- g_{tt} - g_{\phi\phi} \Omega^2}},\\
\Omega & = & \sqrt{\frac{-\partial_r g_{tt}}{\partial_r g_{\phi\phi}}}.\label{eq:omega}
\end{eqnarray}

As previously mentioned, we consider six boson star models: $\Lambda$BS1, $\Lambda$BS2, $\Lambda$BS3, SBS1, SBS2 and SBS3, summarized in Table \ref{tab:solutions}. A noteworthy difference between the $\Lambda$BS and the SBS models stands on the existence of an ISCO. Indeed, the $\Lambda$BS models feature stable circular timelike geodesics for any value of the radial coordinate, whereas the SBS models have an ISCO. The existence of stable circular orbits for $\Lambda$-stars models for all values of the radial coordinate can pose several difficulties when modeling accretion disks around these objects. Indeed, the accreted matter begins to fill each circular orbit up to the surface of the compact object. Over time, and due to the absence of a physical surface in these models, the accreted matter accumulates at the center of the star, potentially leading to the formation of a ``baryonic BH". Nevertheless, as discussed by Torres \cite{Torres:2002td}, due to the slow accretion rates in certain scenarios, even if a BH forms in a timescale of the order of the age of the universe, the mass of such a BH is several orders of magnitude smaller than the mass of the initial boson star. Additionally, the SBS3 model differs from all of the remaining models for having a compacticity large enough to hold a pair of photon spheres, i.e., circular null geodesics.

In this work we assume that the boson star models are of stellar mass. In particular, we adopt a fiducial mass of $M = 14.8 \; M_{\odot}$. This corresponds to the compact object in Cygnus X-1, a galactic X-ray source. This object belongs to a binary system: the companion is a blue supergiant variable star (HDE226868). The stellar wind from the star provides the material for an accretion disk around the X-ray source. The source has been observed in both a low-hard state, dominated by the emission of a hot corona (see for instance \cite{Dove:1997ei,Gierlinski:1996az,Poutanen:1998jc}), and a high-soft state, dominated by the accretion disk, which in this state extends all the way down to the innermost stable orbit. In the latter state, the accretion disk can be modeled as an optically thick and geometrically thin disk, such as those described in Refs. \cite{Novikov:1973kta} and \cite{Page:1974he}. The accretion rate of the source is $\dot{M} \approx 0.1 \dot{M}_{\mathrm{Edd}}$ \cite{Gou:2011nq}, with
\begin{equation}
\dot{M}_{\mathrm{Edd}} = \frac{L_{\mathrm{Edd}}}{c^{2}} \approx 0.2 \times \left(\frac{M}{M_{\odot}}\right) \mathrm{M_{\odot} \; yr^{-1}},
\end{equation}
where $\dot{M}_{\mathrm{Edd}}$ and $L_{\mathrm{Edd}}$ are the Eddington accretion rate and luminosity, respectively. For our particular case, this becomes $\dot{M} \approx 0.3 \times 10^{-8} \; \mathrm{M_{\odot}/yr}$. At this rate, it would take about $10^{8}$ years to accumulate 1 $M_{\odot}$ at the center of the star. Thus, we assume the gravitational influence of this ``baryonic center" on the accreting matter to be negligible with respect to the gravitational influence of the boson star. In view of the remarks above, we consider that the accretion disks extend down to the center of the boson star for the $\Lambda$BS models.

\begin{figure}
\centering
\includegraphics[width=\columnwidth]{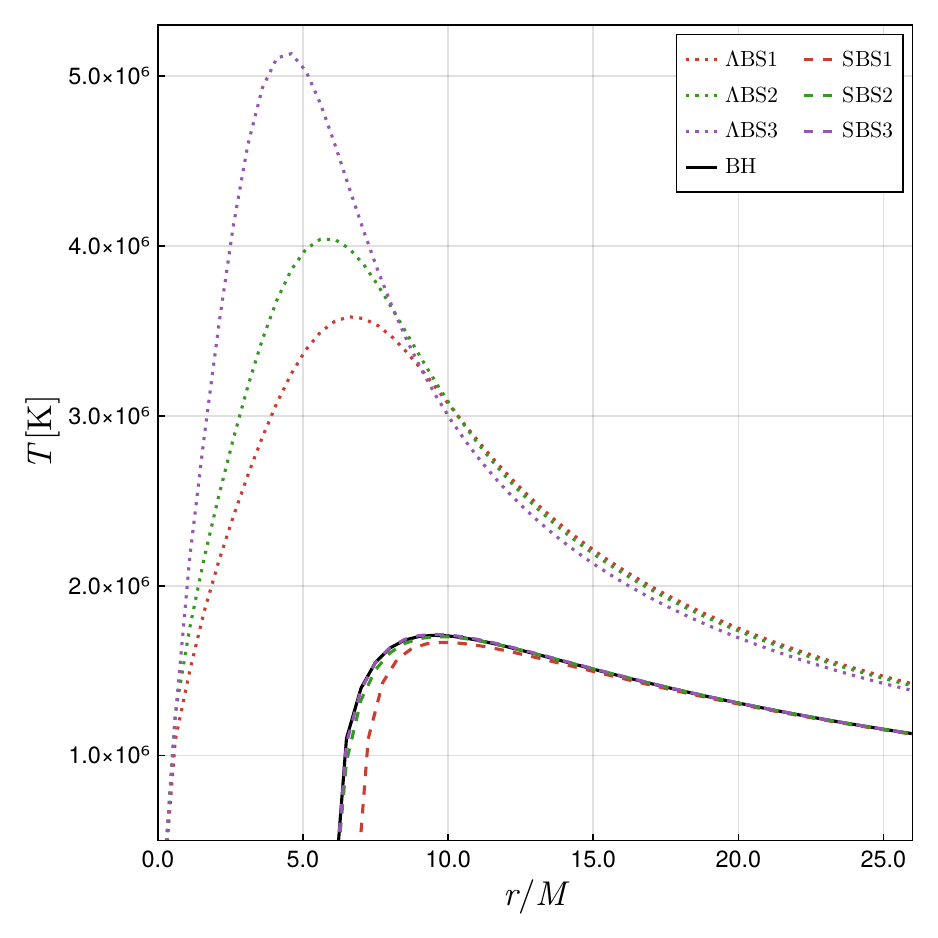}
\caption{Temperature profiles for a relativistic accretion disk around the $\Lambda$BS and SBS models, with a mass of $M = 14.8 \; M_{\odot}$. The thin black line represents the results for the black-hole spacetime with the same mass as a reference.}  
\label{fig:temp}
\end{figure}

If one assumes that the disk is optically thick in the direction orthogonal to the equatorial plane, say the $z$ direction, every surface element radiates as a blackbody with a temperature given by the local effective surface temperature $T(r)$. Through Stefan Boltzmann’s law, we obtain the radial temperature profile
\begin{equation}
T(r) = \left(\frac{Q(r)}{\sigma_{\mathrm{SB}}}\right)^{1/4},
\label{eq:temperature}
\end{equation}
where $\sigma_{\mathrm{SB}}$ denotes the Stefan-Boltzmann constant. In Fig. \ref{fig:temp} we show the temperature profiles as functions of radius for both the $\Lambda$BS and SBS models. For reference, we have also included the corresponding temperature profile for a Schwarzschild black hole case, which we denote the BH scenario. Furthermore, in Table \ref{table:2}, we specify the maximum temperature of the disk and the radius at which the maximum temperature is attained. 


\begin{table}
\centering  
\begin{tabular}{ |p{1cm}||p{1.3cm}|p{1.3cm}|}
 \hline
 Model  &  $R\ \left[r_{\mathrm{g}}\right]$ & $r_{\text{ISCO}}\ \left[r_{\mathrm{g}}\right]$ \\
 \hline
 $\Lambda$BS1   & 9.89994 & -   \\
 $\Lambda$BS2 &   8.62226 & - \\
 $\Lambda$BS3 & 6.77940 & - \\
 SBS1   & 6.58434 & 6.74968\\
 SBS2 &   3.94597 & 6.00000 \\
SBS3 &   3.12060 & 6.00000 \\
\hline
\end{tabular}
\caption{Radius $R$ encapsulating $98\%$ of the mass of the boson star, and radius of the ISCO $r_{\text{ISCO}}$, normalized to the gravitational radius $r_g=GM/c^2$ for the models summarized in Table \ref{tab:solutions}.}
\label{table:1}
\end{table}

Our results indicate that the highest energy fluxes and temperatures are achieved for the $\Lambda$BS models, with more compact stars associated to hotter accretion disks. This is an expected result given that the integral in Eq.~\eqref{flux} extends to $r=0$ in these models, which from Eq.~\eqref{eq:temperature} leads to higher temperatures. On the other hand, due to the presence of an ISCO for the SBS models, the thermodynamic properties of associated to the accretion disks  are more closely related to those of Schwarzschild BHs, including temperature profiles of the same order of magnitude. 

The temperature profiles of the accretion disks of SBS configurations closely resemble those of Schwarzschild. Since the value of the ISCO for these models, and hence the inner edge of the disk, approaches that of Schwarzschild, the most energetic region of the disks in SBS models largely coincides with the black hole spacetime. In particular, the temperature profile of the SBS3 model perfectly mimics that of a Schwarzschild black hole.

\begin{table}
\centering  
\begin{tabular}{ |p{1.5cm}||p{1.7cm} | p{1.5cm}|}
 \hline
 Model  & $T_{\mathrm{max}}$ $\left[\mathrm{K}\right]$ & $r_{\rm max}$ $\left[r_{\mathrm{g}}\right]$ \\
 \hline
 $\Lambda$BS1   & $3.58 \times 10^{6}$ & $6.70$   \\
 $\Lambda$BS2 &   $4.05 \times 10^{6}$ &  $5.84$ \\
 $\Lambda$BS3 & $5.14 \times 10^{6}$ & $4.44$ \\
 SBS1   & $1.67 \times 10^{6}$  & $9.85$\\
 SBS2 &   $1.70 \times 10^{6}$ & $9.72$ \\
SBS3 &   $1.71 \times 10^{6}$ & $9.55$ \\
Schw  BH & $1.71 \times 10^{6}$ & $9.55$ \\
\hline
\end{tabular}
\caption{Maximum temperature $T_{\rm max}$ in CGS units, and radius at which the maximum temperature is achieved $r_{\rm max}$ normalized to the gravitational radius $r_g=GM/c^2$, for the models summarized in Table \ref{tab:solutions}}.
\label{table:2}
\end{table}


\section{Observational properties}\label{sec:results}

To analyze the observational properties of the models previously described, we recur to the open-source Julia-based general-relativistic radiative transfer code \texttt{skylight}\footnote{github.com/joaquinpelle/Skylight.jl} \cite{Pelle:2022sky}. We generate the observed images and spectra that correspond respectively to the thermal continuum and the line emission from accretion disks around boson stars. For this purpose, we incorporated the spacetime metric of Eq.~(\ref{eq:metric}) and its associated Christoffel symbols into \texttt{skylight}, relying on accurate analytical expressions to fit the numerical solutions\footnote{Following what was done in previous publications \cite{Rosa:2022tfv, Rosa:2022toh, Rosa:2023qcv}, we adopted analytical approximations of the numerical solutions with relative errors everywhere smaller than $1\%$, and of the order of $0.01\%$ on average.}. Even though \texttt{skylight} is capable of computing the Christoffel symbols efficiently via automatic differentiation of the metric coefficients, we opted to provide an explicit form of these quantities. This improves the computational speed approximately four-fold in this spherically symmetric scenario, in comparison with the agnostic automatic differentiation version. The code for the production runs and visualizations in this paper using \texttt{skylight} is publicly available\footnote{github.com/joaquinpelle/BosonStars}. The visualizations were created using the package Makie \cite{Danisch:2021Makie}, a high-performance data visualization package for Julia. 

The observation points are located at $r=1000 M$, and at various inclination angles relative to the rotation axis of the disk, $\xi = 5^\circ$, $45^\circ$, $85^\circ$, with a camera resolution of $1200 \times 1200$ pixels for all runs. Angular apertures of the camera varied across models, and the fluxes were measured in the static frame, $t^\mu \propto \partial_t$, along the radial direction. 

As detailed in Section~\ref{sec:disk}, we defined the inner radius of the disk as zero for the $\Lambda$BS models, and $r_\text{ISCO}$ for the SBS models. For all cases, we set the outer radius of the disk at $80M$. For reference, we have also conducted our analysis in Schwarzschild spacetime, referred to as the BH scenario. In this context, the inner radius of the disk is set at the $r_\text{ISCO} = 6M$, with the radial temperature profile calculated as outlined in Section~\ref{sec:disk}. For every configuration considered, the particles composing the disk move in timelike circular geodesics with four-velocity
\begin{equation}
    u^\mu \propto \partial_t + \Omega \partial_\varphi\,,
    \label{eq:fourvelocitydisk}
\end{equation}
where $\Omega$ is the angular velocity given in Eq.~\eqref{eq:omega} and the four-velocity vector is subject to the normalization condition $u_\mu u^\mu = -1$.

\subsection{Thermal continuum emission}

\begin{figure*}
    \centering
    \includegraphics[width = \textwidth]{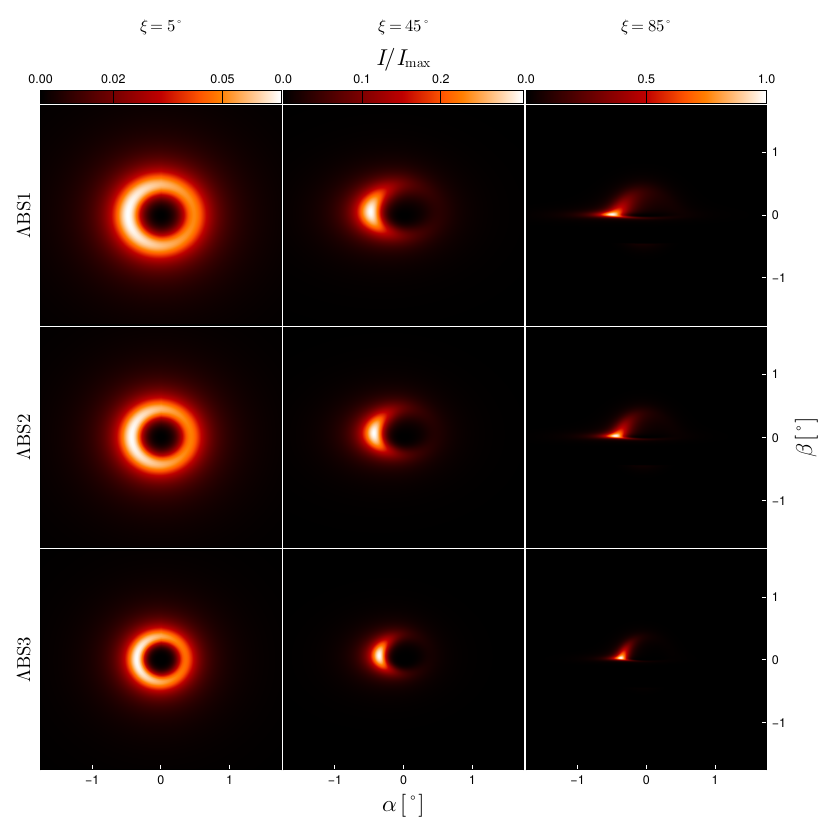}
    \caption{Observed bolometric images showing an accretion disk around a boson star for the $\Lambda$BS models, at inclinations $\xi = 5^\circ$, $45^\circ$, $85^\circ$ with respect to the rotation axis of the disk. The mass of the boson star is $M = 14.8 M_{\odot}$. The observation point is at $r = 1000 \, M$, and the intensity is measured in the staic frame. Here, $\alpha$ and $\beta$ denote the angular coordinates on the observer's sky. The bolometric intensity of the radiation field has been scaled to $I_{\text{max}}= 4.26 \times 10^{22} \, \text{erg} \, \text{cm}^{-2} \, \text{sr}^{-1} \, \text{s}^{-1}$.}
    \label{fig:imagesLBS}
\end{figure*}

\begin{figure*}
    \centering
    \includegraphics[width = \textwidth]{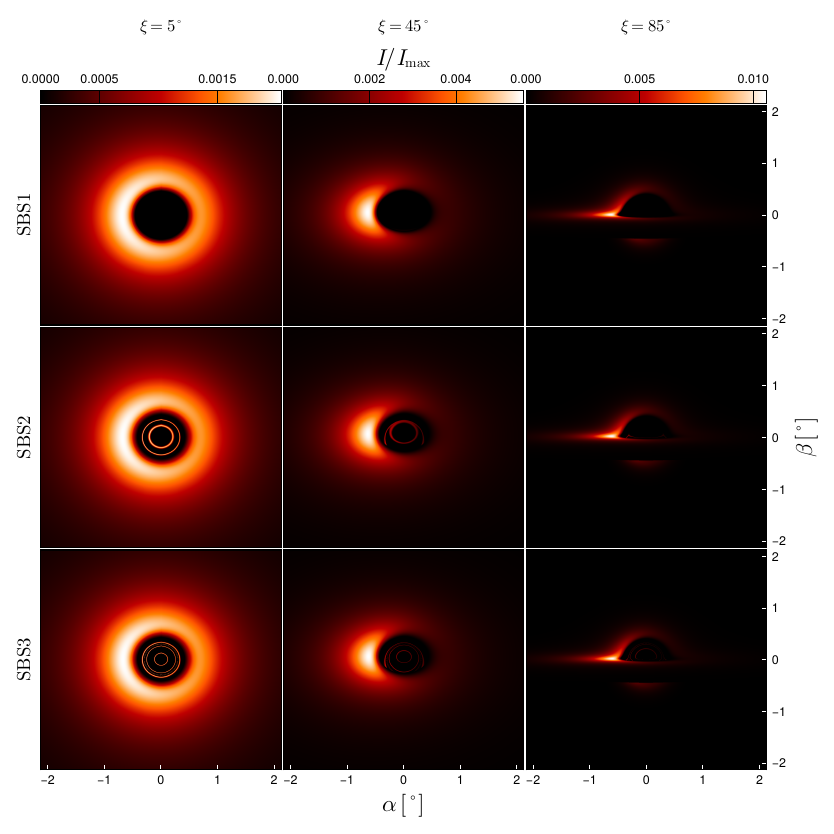}
    \caption{Observed bolometric images showing an accretion disk around a boson star for the SBS models, at inclinations $\xi = 5^\circ$, $45^\circ$, $85^\circ$ with respect to the rotation axis of the disk. The mass of the boson star is $M = 14.8 M_{\odot}$. The observation point is at $r = 1000 \, M$, and the intensity is measured in the staic frame. Here, $\alpha$ and $\beta$ denote the angular coordinates on the observer's sky. The bolometric intensity of the radiation field has been scaled to $I_{\text{max}}= 4.26 \times 10^{22} \, \text{erg} \, \text{cm}^{-2} \, \text{sr}^{-1} \, \text{s}^{-1}$.}
    \label{fig:imagesSBS}
\end{figure*}

\begin{figure*}
    \centering
    \includegraphics[width = \textwidth]{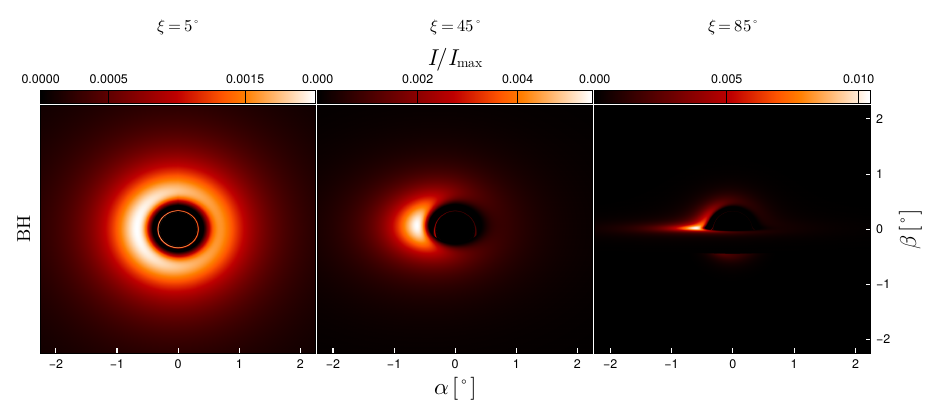}
    \caption{Observed bolometric images showing an accretion disk around a Schwarzschild BH,  at inclinations $\xi = 5^\circ$, $45^\circ$, $85^\circ$ with respect to the rotation axis of the disk. The mass of the BH is $M = 14.8 M_{\odot}$. The observation point is at $r = 1000 \, M$, and the intensity is measured in the static frame. Here, $\alpha$ and $\beta$ denote the angular coordinates on the observer's sky. The bolometric intensity of the radiation field has been scaled to $I_{\text{max}}= 4.26 \times 10^{22} \, \text{erg} \, \text{cm}^{-2} \, \text{sr}^{-1} \, \text{s}^{-1}$.}    
    \label{fig:imagesSCHW}
\end{figure*}

\begin{figure*}
    \centering
    \includegraphics[width = \textwidth]{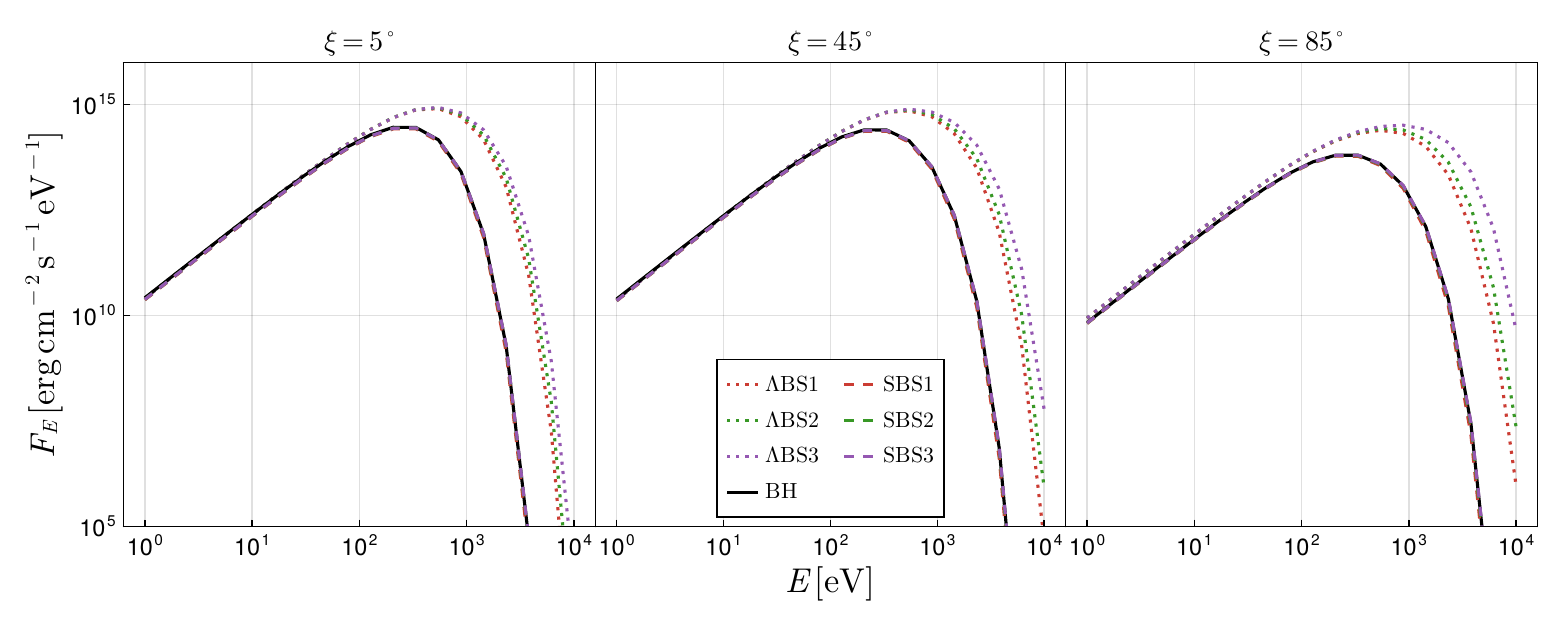}
    \caption{Observed specific fluxes for the thermal continuum emission in the $\Lambda$BS, SBS, and BH models, at inclinations $\xi = 5^\circ$, $45^\circ$, $85^\circ$ with respect to the rotation axis of the disk. The mass of the central object is $M = 14.8 M_{\odot}$. The observation point is at $r = 1000 \, M$, and the flux is measured in the static frame, and along the radial direction.}   
    \label{fig:spectra}
\end{figure*}

Let us start by focusing on the thermal continuum emission from the accretion disk, assuming that each fluid element radiates as a blackbody at the local temperature defined in Eq.~(\ref{eq:temperature}), within its local rest frame. The calculation with \texttt{skylight} takes into account both the local Doppler boosting of the radiation from the rest frame and the gravitational redshift occurring between the source and the observation point. Figures~\ref{fig:imagesLBS}, \ref{fig:imagesSBS}, and \ref{fig:imagesSCHW} depict the bolometric images at various inclination angles ($5^\circ$, $45^\circ$ and $85^\circ$) for the $\Lambda$BS, SBS and BH scenarios, respectively. The associated thermal spectra for these models are presented in Fig.~\ref{fig:spectra}.

One of the most prominent features across the $\Lambda$BS, SBS and BH spacetimes is a darker central region, despite the absence of event horizons in the former two. This can be attributed to two different effects: the drop in the matter temperature towards the center of these configurations, and the stronger gravitational redshift of photons emitted close to the center. This is especially noteworthy in the $\Lambda$BS models, where even though the accretion disks extend down to the center of the star (see Section~\ref{sec:disk}), there is still a shadow-like feature with a similar size to those in the SBS and BH configurations, where the disk is truncated at the ISCO. Notice, however, that the intensity varies greatly between different models, and thus this effect is only visible once the intensity has been normalized. Indeed, as previously mentioned, the $\Lambda$BS disks are hotter than the SBS and BH disks. This difference is reflected in the overall luminosity, with hotter models exhibiting higher luminosity and a thermal spectrum that leans towards higher energies.

Also noteworthy is the effect of radiation beaming at the $85^\circ$ viewing angles, which exhibit higher peak intensities due to the higher negative speeds along the line of sight. Moreover, the $\Lambda$BS and SBS configurations display significant light deflection, particularly at the $85^\circ$ viewing angle. Consequently, the upper and lower sides of the disk portion behind the central object are visible even in an edge-on view, similar to the BH scenario. Furthermore, the SBS models, owing to their compactness, exhibit intricate light-ring structures, whose complexity is made possible by the absence of an event horizon, which allows photons to reach the observer after traversing the interior of the boson star. In particular, the most compact model (SBS3) features four additional light rings in comparison to the BH scenario, one of which is associated with the presence of a pair of photon spheres. The second most compact model (SBS2) features two additional light rings, but neither of them is associated with a photon sphere, since that the latter is absent in this model\footnote{These additional contributions in the SBS2 and SBS3 model that are not associated with the presence of a photon sphere correspond to highly deflected photons which have crossed the equatorial plane once after emission and before reaching the observer.}. In contrast, the mass distributions in the $\Lambda$BS and SBS1 models are not sufficiently compact to deflect light strongly enough to produce light rings.

A closer look at the $\Lambda$BS1, $\Lambda$BS2, and $\Lambda$BS3 models shows a correlation between their size and compactness—the models appear smaller as they become more compact. Interestingly, although SBS configurations are more compact, they appear larger than the $\Lambda$BS models. Recalling that the observed intensities have been normalized, this effect can be attributed to the peak temperature (and hence the most luminous region) lying at larger radii in the SBS configurations.

\subsection{Line emission}

Let us now consider the line emission from the accretion disk illuminated by a corona. The corona, a hotter gas surrounding the accretion disk, contains electrons that can up-scatter thermal photons from the disk to higher energies. Part of these photons illuminate back the disk and induce line emissions such as iron lines, depending on the ion abundances of the disk. In the local emission rest frame, the spectral line is narrow, but the differential effects of gravitational redshift and Doppler boosting from various parts of the disk broaden the line in the reference frame of a distant observer. The shape of the observed line also depends on the emissivity profile on the disk, which depends on the illumination pattern.

The geometry of the corona, which determines the emissivity profile $I_e$, remains unknown even in the standard black hole scenario. Previous works have considered ad-hoc emissivity profiles, such as that produced in the lamppost geometry \cite{cao2016, lu2003} where the corona is represented as a single point at a height $h$ along the rotational axis of the accretion disk. In flat spacetime, this results in
\begin{equation}
    I_e \propto \frac{h}{\left(h^2+r^2\right)^{3/2}} \,,
    \label{eq:lamppost-flat}
\end{equation}
In this work, we explore an improvement of this setup by incorporating the relativistic effects of the boson star spacetimes, following the methodology previously applied for black holes in Ref. \cite{dauser2013}. The point-like corona is assumed to emit isotropically, and we model this by taking a bundle of $5 \times 10^6$ photons starting at $r=h$ along the polar axis, with spatial momenta isotropically distributed in the static frame. We trace the corresponding null geodesics with \texttt{skylight} and collect those that intersect the accretion disk. We take $50$ radii, $r_i$, between the inner and outer radii of the accretion disk, and we bin the geodesics into small annuli delimited by the $r_i$. The area $A_i$ of each annulus in the coordinate frame is, approximately,
\begin{equation}
    A_i \approx 2\pi \sqrt{g_{rr}g_{\varphi \varphi}} \Delta r_i\,,
    \label{eq:area-bin}
\end{equation}
where the metric is evaluated at $r_i$, and $\Delta r_i$ is the radial extent of the annulus. The corresponding area in the rest frame is obtained by multiplying Eq.~\,(\ref{eq:area-bin}) by the Lorentz factor of the disk particles, $\gamma = u^t$, where $u^\mu$ is given in Eq.~\,(\ref{eq:fourvelocitydisk}) with suitable normalization. The number density of the rays then becomes
\begin{equation}
    n_i = \frac{\mathcal{N}_i}{A_i \gamma_i}\,,
    \label{eq:raynumberdensity}
\end{equation}
where $\mathcal{N}_i$ is the number of rays per radial bin. 

\begin{figure*}
    \centering
    \includegraphics[width = \textwidth]{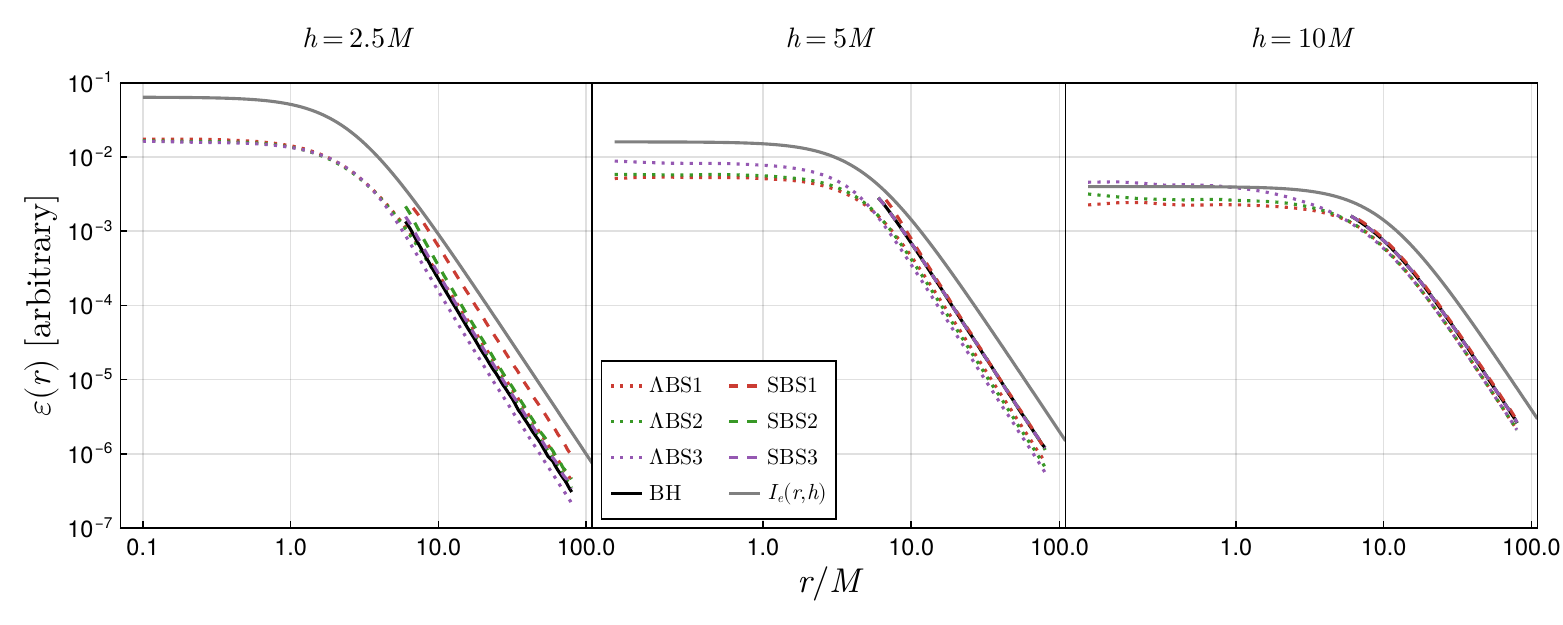}\\
    \includegraphics[width = \textwidth]{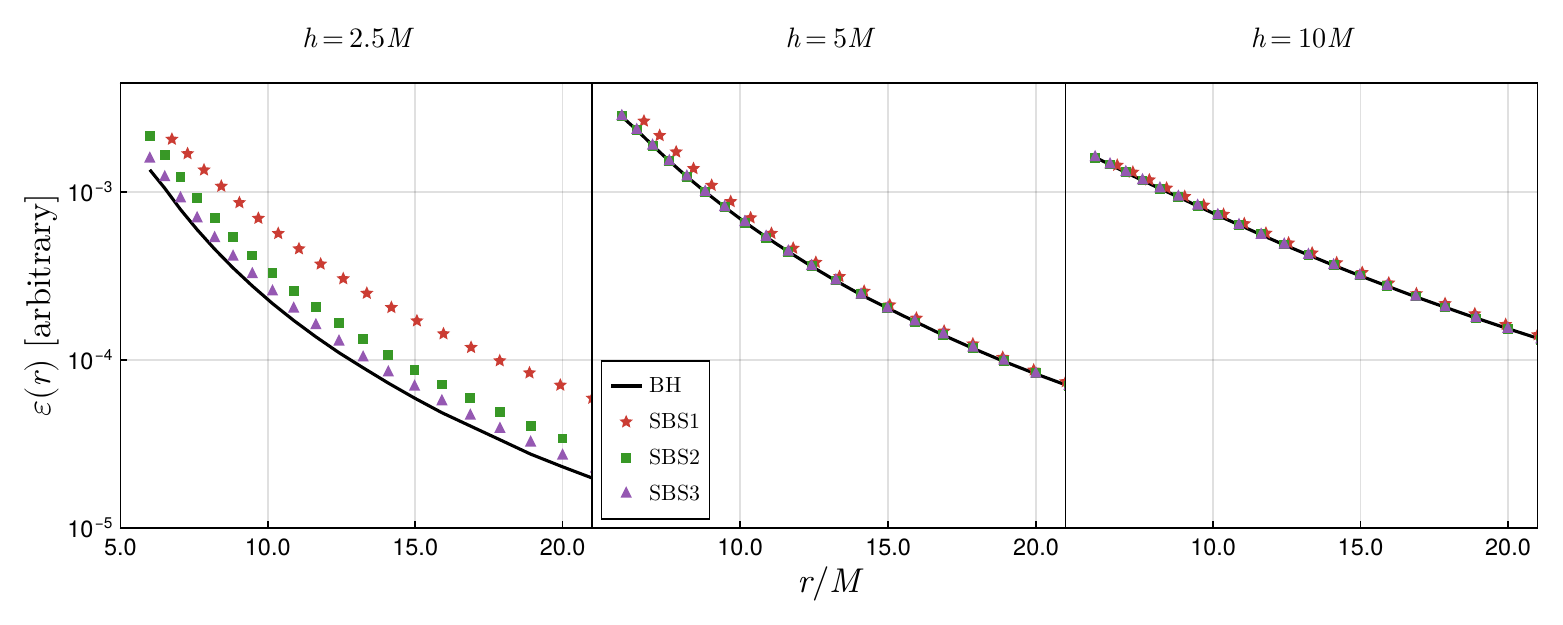}
    \caption{Emissivity profiles in the lamppost corona model for the $\Lambda$BS, SBS, and BH scenarios. Top row: the profiles are presented for various lamppost heights, alongside the corresponding flat-spacetime prescriptions from Eq.(\ref{eq:lamppost-flat}); bottom row: zoomed view of the same emissivity profiles for the SBS and BH scenarios.}
    \label{fig:profiles}
\end{figure*}

On the other hand, we approximate the photon flux at the corona as a power law of the form
\begin{equation}
dN = K E_0^{-\Gamma} dt_0 dE_{0} \, ,
\end{equation}
where $K$ is a constant, $\Gamma$ is the photon index, $E$ is the energy of the emission line, and the zero subscript refers to quantities evaluated at the corona. Since the photon number is conserved along each ray, at the accretion disk we must have 
\begin{equation}
dN = K g^{\Gamma} E^{-\Gamma}  dt dE \, ,
\end{equation}
where $g = E/E_0$ is the redshift factor between the corona and the accretion disk, and we have used the fact that $dE/dE_0 = dt_0/dt = g$. Therefore, the energy density illuminating the disk is
\begin{equation}
d\mathcal{E} = E n dN = K g^\Gamma E^{-\Gamma+1} n dt dE\,,
\end{equation} 
and the radial emissivity profile can be written as
\begin{equation}
\varepsilon \propto \frac{d\mathcal{E}}{dt dE} = K g^\Gamma E^{-\Gamma+1} n \,.
\end{equation}
The emissivity profiles for the $\Lambda$BS, SBS and BH scenarios, corresponding to three different lamppost heights, are presented in Fig.\,\ref{fig:profiles}, together with the flat-spacetime analytical profile from Eq.(\ref{eq:lamppost-flat}). Interestingly, the profiles are similar across different scenarios, and they appear to align with the flat-spacetime prescriptions. However, this resemblance should not lead to the interpretation that flat spacetime is a suitable approximation. To illustrate this point, in Fig.\,\ref{fig:profiles} we provide a zoomed view of the emissivity profiles for the SBS and BH scenarios close to the inner edge of the disks. This inner region, where the resemblance is less accurate, contributes most significantly to the observed lines. Even moderate differences in this critical area can strongly influence the resulting line. We clarify these points in the analysis that follows.

In the $\Lambda$BS scenarios, where the inner radius is zero, the emissivity profile flattens at small radii. This characteristic amplifies the contribution of the central region to the spectrum, leading to line shapes that are heavily concentrated at lower energies due to the strong redshift of photons coming from that region. Since we regard this situation as unphysical and lack a physically grounded criterion to truncate the central region, we have chosen to exclude these scenarios from our subsequent considerations.

\begin{figure*}
    \centering
    \includegraphics[width = \textwidth]{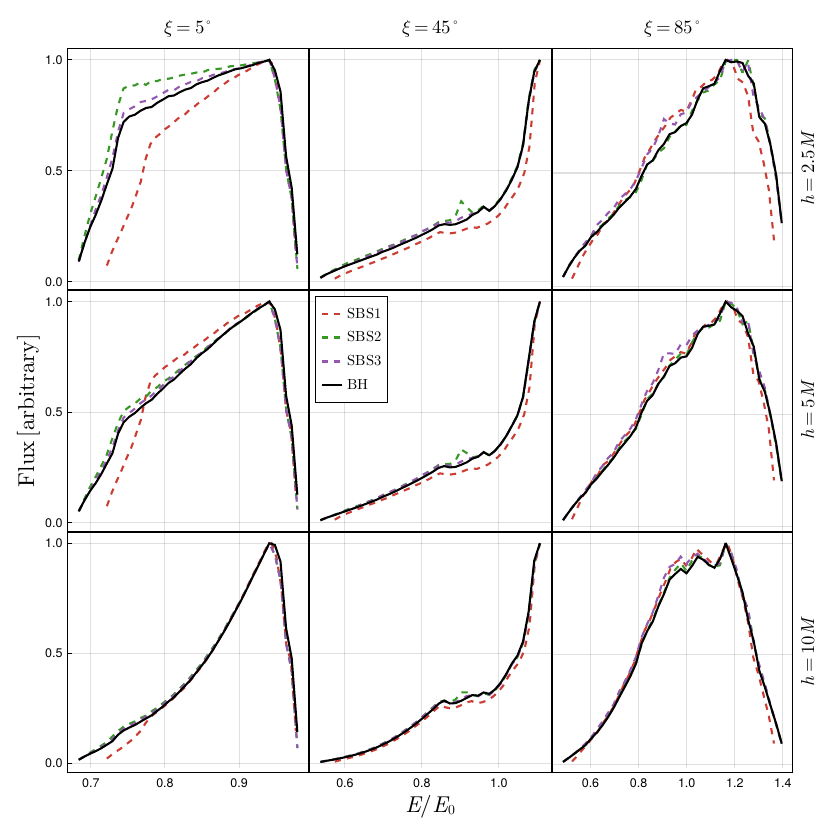}
    \caption{Observed line shapes in the lamppost corona model for the SBS and BH scenarios, for various lamppost heights and observer inclination angles. $E_0$ is the energy of the emission line in the rest frame of the accretion disk.}   
    \label{fig:lines}
\end{figure*}

Turning our attention to the SBS and BH scenarios, the observed line shapes are presented in Fig.\,\ref{fig:lines} for various observer inclination angles. We note that, as mentioned before, the shapes can differ significantly even if the emissivity profiles follow similar patterns. This is clear in the line shapes for $\xi = 5^{\circ}$, and $h=2.5M$, $5M$. However, some other line shapes in the SBS configurations do resemble the BH scenario closely, as for $h=5M$, $10M$, and $\xi=45^{\circ}$, $85^{\circ}$. This suggests that, under certain conditions e.g. relatively large values of $h$ and $\xi$, it might be difficult to distinguish a compact boson star from a non-spinning BH based solely on spectroscopy of emission lines. These results emphasize how exotic compact objects with large compacticities, particularly large enough to produce an ISCO at a radius similar to that of a BH, can display observational features similar to those of BHs and effectively behave as BH mimickers.

\section{Conclusions}\label{sec:concl}

In this work, we have analyzed the observational properties of scalar boson star models featuring fourth and sixth-order scalar field self-interactions, in an astrophysical setting in which they are surrounded by an optically thick and geometrically thin accretion disk described by the Novikov-Thorne model. In particular, we have used the Julia-based radiative transfer software \texttt{skylight} to produce the observed images of the disks and calculate the relativistic broadening of emission lines, including a lamppost corona model with fully relativistic considerations.

For boson stars with quartic self-interactions, dubbed as $\Lambda$BS models, we have verified that the observational properties of these configurations differ strongly from the BH case. Indeed, because these models do not possess an ISCO and, consequently, the accretion disks extend to the central region, the temperature profiles of these configurations attain larger values in comparison to those of the BH scenario. Nevertheless, it is noteworthy that the effects of the gravitational redshift and lower temperatures in the central region lead to the appearance of a ring-like shadow feature in the observer's screen anyway. Our results thus indicate that these models are unsuitable to fit current observations.

As for the SBS models, there is a closer similarity in the observed properties compared to the BH scenario. In fact, the temperature profiles obtained for these models more closely resemble those of the Schwarzschild black hole; in particular, we found that the temperature profile of the accretion disk in SBS3 models is almost identical to that of the BH; the effects of light deflection lead to similar observed shadows, with the SBS models featuring additional light-ring contributions due to the absence of an event horizon; the broadening effects on emission lines are qualitatively comparable, especially for large heights of the corona. These results emphasize SBS models as suitable candidates for BH mimickers.

Even though the SBS models, especially SBS3, share many observational features with the BH scenario, they might have limitations from an astrophysical point of view. Indeed, the models considered are static and spherically symmetric, whereas one expects astrophysical objects to exhibit rotation. Rotating bosonic star configurations have been found in the literature \cite{Brito:2015pxa,Liebling:2012fv}, and thus it would be interesting to analyze these configurations and compare the resulting observational properties with those of the Kerr scenario. Fully understanding this comparison, however, would require a more systematic exploration, a task we reserve for future work.

\begin{acknowledgments}
J.L.R. acknowledges the European Regional Development Fund and the programme Mobilitas Pluss for financial support through Project No.~MOBJD647, project No.~2021/43/P/ST2/02141 co-funded by the Polish National Science Centre and the European Union Framework Programme for Research and Innovation Horizon 2020 under the Marie Sklodowska-Curie grant agreement No. 94533, Fundação para a Ciência e Tecnologia through project number PTDC/FIS-AST/7002/2020, and Ministerio de Ciencia, Innovación y Universidades (Spain), through grant No. PID2022-138607NB-I00. J.P. acknowledges financial support from CONICET under a doctoral research fellowship. D. P. acknowledges the support from CONICET under Grant No. PIP 0554 and AGENCIA I+D+i under Grant PICT-2021-I-INVI-00387.

This work used computational resources from CCAD – Universidad Nacional de Córdoba (https://ccad.unc.edu.ar/), which is part of SNCAD – MinCyT, República Argentina.
\end{acknowledgments}

\section*{Author contributions}
J. L. R. implemented the numerical procedure to obtain the boson star configurations and developed the corresponding analytical approximations to be implemented in the radiative transfer code. D.P. constructed the accretion disk models and computed the corresponding heat fluxes and temperature profiles. J.P. implemented the boson star spacetimes and the accretion disk model in the radiative transfer code, conducted the simulations, and developed the figures. All authors participated in discussing the results and contributed significantly to writing the manuscript.



\end{document}